# SIMULATION OF VALVELESS MICROPUMP AND MODE ANALYSIS

*W. P. Lan, J. S. Chang, K. C. Wu, and Y. C. Shih*

Institute of Applied Mechanics, National Taiwan University, Taiwan
([†]Corresponding author Email: r93543057@ntu.edu.tw)

## Abstract

In this work, a 3-D simulation is performed to study for the solid-fluid coupling effect driven by piezoelectric materials and utilizes asymmetric obstacles to control the flow direction. The result of simulation is also verified. For a micropump, it is crucial to find the optimal working frequency which produce maximum net flow rate. The PZT plate vibrates under the first mode, which is symmetric. Adjusting the working frequency, the maximum flow rate can be obtained. For the micrpump we studied, the optimal working frequency is 3.2K Hz. At higher working frequency, say 20K Hz, the fluid-solid membrane may come out a intermediate mode, which is different from the first mode and the second mode. It is observed that the center of the mode drifts. Meanwhile, the result shows that a phase shift lagging when the excitation force exists in the vibration response. Finally, at even higher working frequency, say 30K Hz, a second vibration mode is observed.

Keywords: solid-fluid coupling, valveless micropump, piezoelectrical, mode analysis

## 1. INTRODUCTION

Since the inception of micro-electro-mechanical system (MEMS), piezoelectric materials have played a crucial role of the smart materials and structures technology. With the advancement of technology, microfluidic devices can be fabricated by MEMS. Piezoelectric actuators have been widely used in the valveless micropump, because the valveless micropumps actuated by piezoelectric plates make the device smaller than the others and easy for use.

Fan *et al* [1] obtained distinct net flow rates for different working frequencies. They also showed that different modes of interaction between fluid and structure made the flow rate change. Yang *et al* [2] used the asymmetric obstacles controlling the direction of the flow and volume flow rate. They derived the mathematic formula which substitutes as the vibration of piezoelectric plate in order to simplifying simulation. The simulation includes the dimension of the obstacle and the viscous effect of flow rate. Pan *et al* [3] showed that there were the essential relations among inertial, viscosity force and the dimension of the micropumps, for the high working frequencies, the order of the magnitudes of inertial force and viscous force were very close. They show that this phenomenon had no concern with fluid and the working frequencies. In addition, the phase lag phenomenon occurring at the interface between the fluid and structure was observed in this work.

This research is about the performance of piezoelectric actuated valveless micropump with consideration of the three-way electric-mechanical-fluid couplings. We can obtain the information about the effect of fluid damping in micropump from many references. In previous research, they neglected the effect of damping. They usually assumed that when power supply is turned on, the piezoelectric plate will have a displacement in the air. In this way, the decayed phenomenon of water driven by vibration of piezoelectric film cannot be estimated. In our opinion, we can find another method to combine displacement with damping effect.

## 2. THEORY

Figure 1 and 2 show the schematic diagram of our PZT actuated valveless pump. The pump is fabricated form silicon by MEMS. Two asymmetric obstacles were put in the flow channel to control the net flow direction. On the circular chamber is covered one side PZT film. The PZT plate is driven by ac voltage, with the frequency ranged from 2K to 30K Hz. Fluid used in this study is water. The material properties used and the dimension of the pump are listed in Table 1 to 3, and Table 4, respectively. In this work, we do a complete 3D simulation, considering the interaction between solid and fluid.

### 2.1 The flow field

We assume that the density $\rho$ and viscosity $\eta$ of the modeled fluid are constant, not be affected by temperature and concentration.

The governing equations of continuity and three-dimensional momentum can be expressed as follows:

$$\nabla \cdot \vec{V} = 0 \quad (1)$$

$$\rho \frac{\partial u}{\partial t} + \rho \vec{V} \cdot \nabla u - \eta \nabla^2 u + \frac{\partial p}{\partial x} = F_{E,x} \quad (2)$$

$$\rho \frac{\partial v}{\partial t} + \rho \vec{V} \cdot \nabla v - \eta \nabla^2 v + \frac{\partial p}{\partial y} = F_{E,y} \quad (3)$$

$$\rho \frac{\partial w}{\partial t} + \rho \vec{V} \cdot \nabla w - \eta \nabla^2 w + \frac{\partial p}{\partial z} = F_{E,z} \quad (4)$$

where $u$, $v$, $w$ are the velocity components of $\vec{V}$, $\eta$ is the dynamic viscosity of the fluid, $\rho$ is the density of the fluid, and $p$ is pressure.

### 2.2 The structure field

For the general elastic materials, the equation of motion, the relationship between the strains and the displacements and charge equation of electrostatics, are:

$$\sigma_{ij,j} + f_i = \rho \ddot{u}_i \quad (5)$$

$$\varepsilon_{ij} = \frac{1}{2}\left(u_{i,j} + u_{j,i}\right) \quad (6)$$

$$D_{i,i} = 0 \quad (7)$$

$$E_i = -\phi_{,i} \quad (8)$$

where the $\sigma_{ij}$ is the stress tensor, $f_i$ is the body force, $\rho$ is the density, $u_i$ is the displacement vector, $\varepsilon_{ij}$ is the strain tensor, $D_i$ is the electric displacement vector, $E_i$ is the electric field, $\phi$ is the electrical potential.

The linear constitutive equations for homogeneous piezoelectric crystals are:

$$\sigma_{ij} = C^E_{ijkl}\varepsilon_{kl} - e_{kij}E_k \quad (9)$$

$$D_i = e_{ikl}\varepsilon_{kl} + \varepsilon^s_{ik}E_k \quad (10)$$

where $C^E_{ijkl}$ is the elastic material constants; $\varepsilon^s_{ik}$ is the dielectric constants, the $e_{kij}$ is the piezoelectric stress constants.

### 2.3 Boundary conditions

We assume the interface between fluid and the piezoelectric plate to be a clamped plate. The displacement, velocity, and curvature of the surrounding of the plate should be zero. The boundary conditions of the clamped plate can be expressed as:

$W = 0 \quad x, y \in \partial\Omega \ (at\ edge)$

$\dfrac{\partial^2 W}{\partial x^2} = \dfrac{\partial^2 W}{\partial y^2} = 0 \quad x, y \in \partial\Omega$

$\dfrac{\partial W}{\partial t} = 0 \quad x, y \in \partial\Omega$

**Table 1:** The fluid property

| The depth of microchannel | 80 $\mu m$ |
|---|---|
| Density | 1000 $Kg/m^3$ |
| Kinematic constant | 1.1E-6 $m^2/s$ |

**Table 2:** The property of glass

| Thickness | 500 $\mu m$ |
|---|---|
| Density | 2330 $Kg/m^3$ |
| Material constants | $[C_{ijkl}] = \begin{bmatrix} 1.65 & 0.63 & 0.63 & 0 & 0 & 0 \\ 0.63 & 1.65 & 0.63 & 0 & 0 & 0 \\ 0.63 & 0.63 & 1.65 & 0 & 0 & 0 \\ 0 & 0 & 0 & 0.79 & 0 & 0 \\ 0 & 0 & 0 & 0 & 0.79 & 0 \\ 0 & 0 & 0 & 0 & 0 & 0.79 \end{bmatrix} 10^{11} N/m^2$ |

**Table 3:** The property of piezoelectric plate

| Thickness | 200 $\mu m$ |
|---|---|
| Density | 2330 $Kg/m^3$ |
| Material constants | $[C_{ijkl}] = \begin{bmatrix} 12.1 & 7.54 & 7.52 & 0 & 0 & 0 \\ 7.54 & 12.1 & 7.52 & 0 & 0 & 0 \\ 7.52 & 7.52 & 11.1 & 0 & 0 & 0 \\ 0 & 0 & 0 & 2.11 & 0 & 0 \\ 0 & 0 & 0 & 0 & 2.11 & 0 \\ 0 & 0 & 0 & 0 & 0 & 2.28 \end{bmatrix} 10^{10} N/m^2$ |
| Piezoelectric stress constants | $[E_{ijkl}] = \begin{bmatrix} 0 & 0 & 0 & 0 & 12.3 & 0 \\ 0 & 0 & 0 & 12.3 & 0 & 0 \\ -5.4 & -5.4 & 15.8 & 0 & 0 & 0 \end{bmatrix} C/m^2$ |

*PZT-5A

**Table 4:** The size dimension of micropump

| W | W1 | L | L1 | $\alpha$ | r |
|---|---|---|---|---|---|
| um | um | um | um | | um |
| 522 | 40 | 1093 | 1000 | $7^0$ | 3000 |

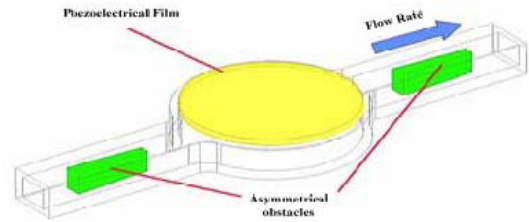

**Figure 1:** Diagram of micropump

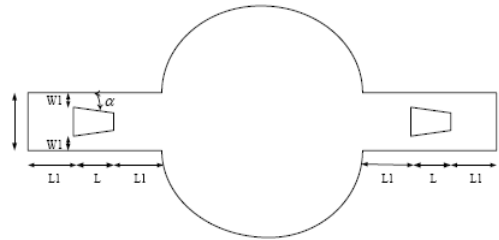

**Figure 2:** Top view of the micropump

## 3. SIMULATION RESULTS

The natural frequency analysis about the structure made by PZT and glass are presented in Figure 3. The first, second, and third natural frequencies of the solid structure are 62812, 168124, and 289100 Hz respectively.

The most significant purpose of the micropump is the relationship between the flow rate and working frequency. In this paper, we applied the fixed voltage and varied the frequency to find the maximum flow rate. Figure 4 is the simulation result of which the fixed voltage is 40 volt with testing input with sinusoidal signal (80 Vp-p, Variation frequency), and the fluid is water. The optimal frequency of simulation is 3200 Hz and the experiment is 3100 Hz. And their corresponding flow rates 157.56 $\mu l/min$ and 155.58 $\mu l/min$, respectively.

Another factor affecting flow rate directly is the mode shape change. Under the first mode, we can find the curvature varying insignificantly. To avoid producing resonance at the interface, we should let the mode shape stay on the first mode by controlling the working frequency. If the mode shape exhibits the second mode, the efficiency of pumping rate may be decreased. Figure 5 and Figure 6 show the mode at the interface when the piezoelectric plate was applied 40 volt with working frequency of 20 K and 30 K Hz.

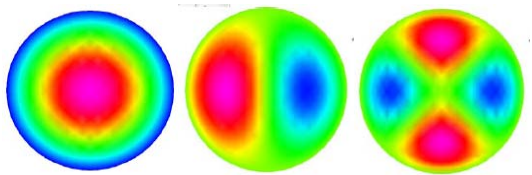

**Figure 3:** From left diagram to right diagram, the shape of the nature frequency analysis of the PZT and glass. The 1st nature frequencies to the 3rd are 62812 Hz, 168124 Hz, and 289100 Hz

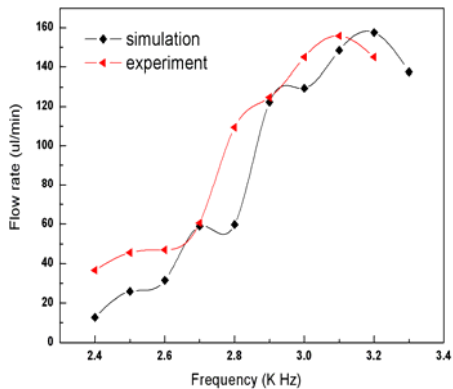

**Figure 4:** The flow rate of water at different working frequency.

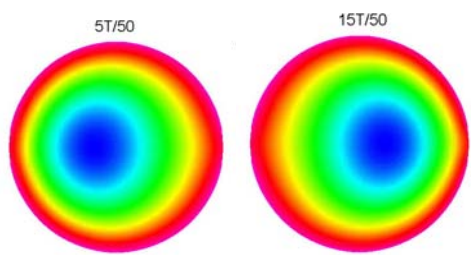

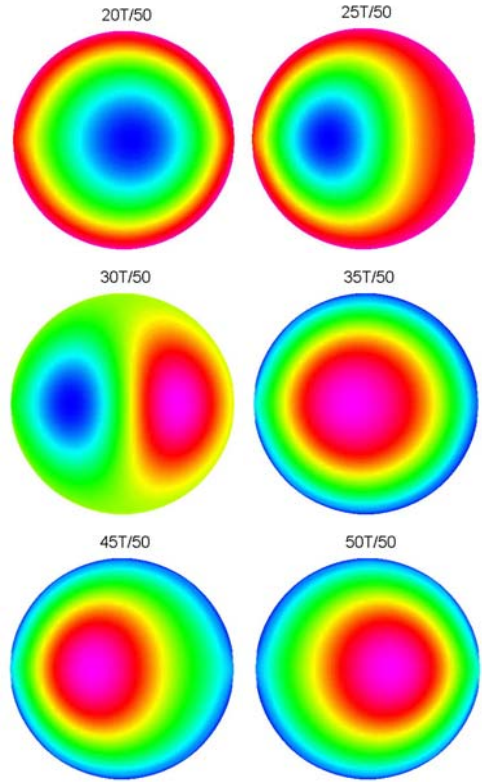

**Figure 5:** The variation of color represent the mode shape change at the interface between fluid and structure with piezoelectric plate is applied 40 voltage and working frequency of 20 K Hz.

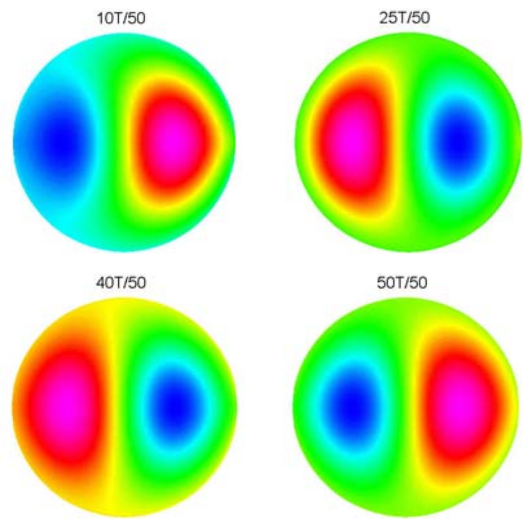

**Figure 6:** The variation of color represent the mode shape change at the interface between fluid and structure with piezoelectric plate is applied 40 voltage and working frequency of 30 K Hz.

## 4. CONCLUSIONS

This 3D simulation presents the solid-fluid coupling study of the micropump actuated by piezoelectric plate with variant working frequency. The results show the relationship between frequency and flow rate. It is unusual that if the working frequency increases, the flow rate may not increase. So, the optimal analysis is obvious essential for micropump. The mode shape exhibits the first mode in the optimal frequency for the size of piezoelectric plate that we chose. If we choose the other sizes of piezoelectric plate are chosen, the optimal frequency may appear at higher modes. Further study is undergoing.

## 5. ACKNOWLEDGEMENTS

Financial support from National Science Council under the Grant No. NSC 95-2120-M-002-006 is greatly appreciated. We are also grateful to the National Center for High-performance Computing for computational support.